\documentstyle[12pt,epsfig]{article}
\def\setb@se#1{\baselineskip=#1 \normalbaselineskip=#1}
\setb@se{14pt}
\newcommand{\labelcaption}[2]{\caption[#1]{\label{#1}#2}}
\newcommand{\be}{\begin{equation}}
\newcommand{\ee}{\end{equation}}
\newcommand{\bea}{\begin{eqnarray}}
\newcommand{\eea}{\end{eqnarray}}
\newcommand{\baf}{{\bar f}}
\newcommand{\bet}{{\bar\eta}}
\newcommand{\ov}[2]{{#1\over #2}}
\newcommand{\rh}{r_{\rm h}}
\newcommand{\fh}{{\bar f}_{\rm h}}
\newcommand{\mh}{\mu_{\rm h}}
\newcommand{\ph}{\psi_{\rm h}}
\newcommand{\etam}{\eta_{\rm max}}
\newcommand{\betam}{\bet_{\rm max}}

\begin{document}

\begin{titlepage}
\hbox to\hsize{%

  \vbox{%
        \hbox{MPI-PhT/99-31}%
%
%
        \hbox{LIU-SC-TCSG-7/1999005}%
%
%
        \hbox{NSF-ITP-99-088}%
        \hbox{August 1999}%
       }}

\vspace{3 cm}

\begin{center}
\Large{Some Remarks on Gravitational Global Monopoles}

\vskip5mm
\large
Dieter Maison

\vspace{3mm}
{\small\sl
Max-Planck-Institut f\"ur Physik\\
--- Werner Heisenberg Institut ---\\
F\"ohringer Ring 6\\
80805 Munich (Fed. Rep. Germany)\\}

\vspace{2mm}

Steven L. Liebling\\

\vspace{3mm}
{\small\sl
Theoretical and Computational Studies Group\\
     Southampton College, Long Island University\\
     Southampton, NY 11968 USA\\}

\end{center} 
\vspace{20 mm}
\begingroup \addtolength{\leftskip}{1cm} \addtolength{\rightskip}{1cm}

\begin{center}\bf Abstract\end{center}
\vspace{1mm}\noindent
Using mainly analytical arguments, we derive the exact relation
$\etam=\sqrt{\ov{3}{8\pi}}$ for the maximal vacuum value of the Higgs field
for static gravitational global monopoles. For this value, the global
monopole bifurcates with the de Sitter solution obtained for vanishing
Higgs field. In addition, we analyze the stability properties of the
solutions.
\endgroup
\end{titlepage}
\newpage
In a recent publication \cite{Liebling}, one of us (SL) presented numerical
results on static, self-gravitating global monopoles.
Considered as topological defects, global monopoles have been discussed
in connection with `topological inflation' \cite{Linde, Vilenkin}.
More recently perturbations of static global monopoles were studied in
the context of `critical behaviour' in spherically symmetric gravitational
collapse \cite{LieblingC}.

In the present paper, we shall derive some of the qualitative and quantitative
results of \cite{Liebling} from the field equations -- a set of non-linear
ODEs -- obeyed by these solutions.
As described in \cite{Liebling} there is a one-parameter family of static
global monopoles parametrized by the Higgs vacuum value $\eta$ (taken in units of
$M_{\rm Pl}=1/\sqrt{G}$) for $0\leq\eta<\etam\approx \sqrt{3/8\pi}$.
As long as $\eta<\sqrt{1/8\pi}$ the space-time of these solutions has a
deficit solid angle at infinity becoming equal to $4\pi$ for
$\eta=\sqrt{1/8\pi}$.
Solutions with $\sqrt{1/8\pi}<\eta<\etam$ were found to have a cosmological
horizon, outside of which the solutions oscillate with decreasing amplitude
about their asymptotic values.
As will be demonstrated, this behaviour can be readily understood from the 
field equations.

Our main result is a simple derivation of the value $\etam$ as being exactly
equal to $\sqrt{3/8\pi}$.
At this value of $\eta$ the numerically determined global monopoles are
found to bifurcate with the de Sitter solution obtained for vanishing Higgs
field -- the non-vanishing vacuum energy yielding a cosmological constant.

A linear stability analysis of both types of solutions shows the stability
of the global monopole for all values $0\leq\eta<\etam$, whereas the
de Sitter solution (considered as a solution of the Einstein-Higgs system) 
is stable against perturbations with support inside the horizon 
for $\eta>\sqrt{3/8\pi}$ and unstable for 
smaller values of $\eta$ with an accumulation of unstable modes for 
$\eta\to 0$.  
The stability of the global monopole is in agreement with
numerical studies of time-dependent solutions performed in 
\cite{LieblingC}.
\begin{figure}   
\epsfig{file=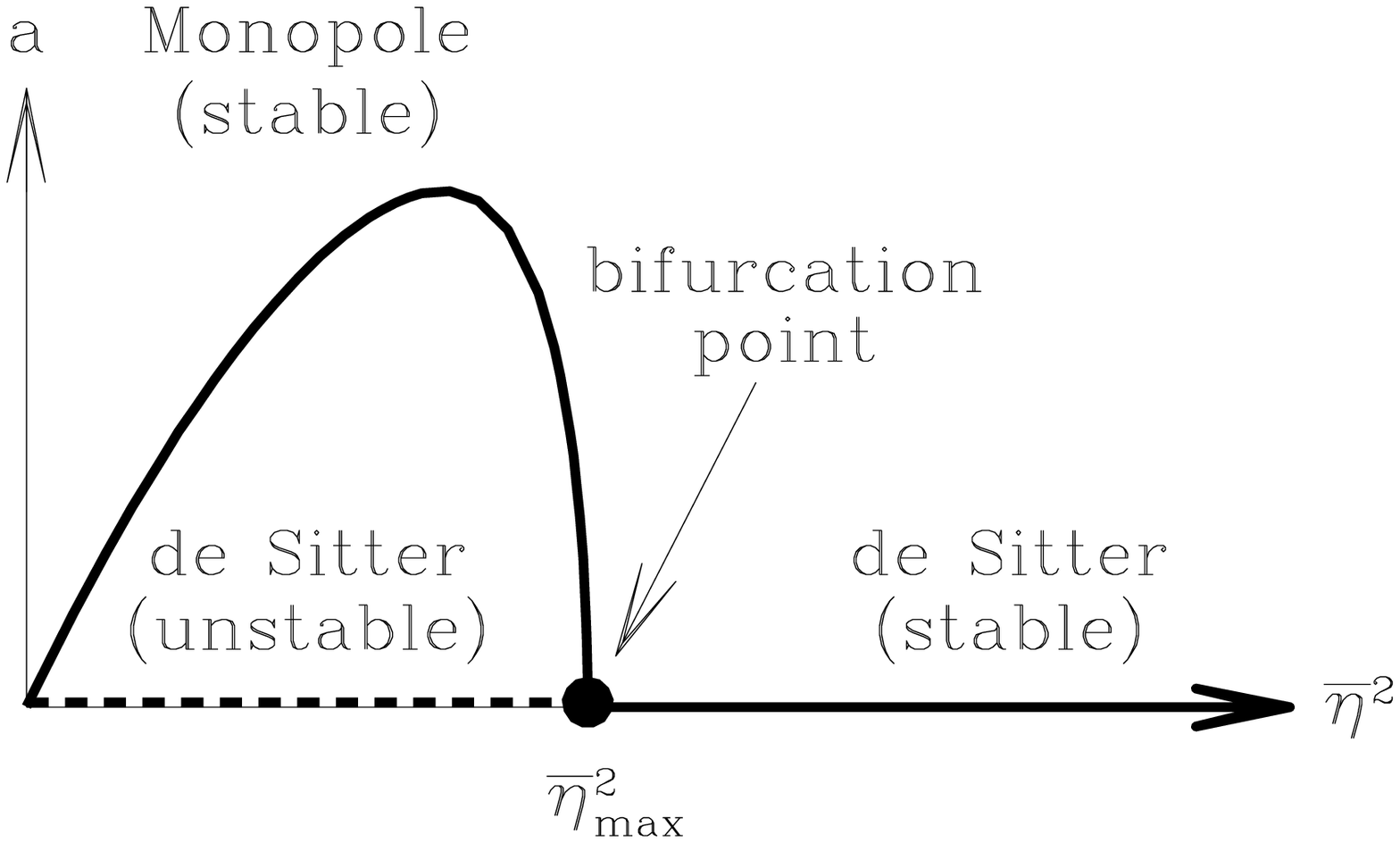,width=9cm,height=9cm}
\vspace{-1.5in}
\labelcaption{figz}{Schematic of the space of static solutions.
For large $\bet^2$, only de Sitter exists and is stable. As
$\bet^2$ decreases, a bifurcation occurs and de Sitter exchanges
stability with the static monopoles.
}
\end{figure}

Following the notation of \cite{Liebling}, we put $\phi^a=f(r)\hat r^a$
for the Higgs field and     
\be                                                  \label{metric}
ds^2=-A^2\mu~dt^2+\ov{dr^2}\mu+r^2d\Omega^2
\ee
for the spherically symmetric line element in Schwarzschild coordinates.
The resulting static field equations are (we prefer to use rationalized
units putting $\baf=\sqrt{4\pi}f$, $\bet=\sqrt{4\pi}\eta$, $\lambda=2\pi$ as compared to
$f$ and $\eta$ from \cite{Liebling}) 
\bea\label{fequs}                                                          
\baf'&=&\psi  \label{fequs_f}\\
\psi'&=&\ov{\baf}{r^2\mu}\Bigl[2+\frac{r^2}{2}(\baf^2-\bet^2)\Bigr]-
         \psi\Bigl[\ov{2}{r}+r\psi^2+\ov{\mu'}{\mu}\Bigr]\\ 
\mu'&=&\ov{1-\mu}{r}-r\mu\psi^2-\ov{2\baf^2}{r}-
        \ov{r}{4}\bigl(\baf^2-\bet^2\bigr)^2 \label{fequs_mu}\\
A'&=&r\psi^2 A \label{fequs_a}.
\eea
Solutions with a regular origin obey the boundary conditions
\be                                                  \label{bdcon}
\baf(r)=ar+O(r^3),\quad \psi(r)=a+O(r^2),\quad \mu(r)=1+O(r^2),\quad
A(r)=A_0+O(r^2)
\ee
and are uniquely specified by the choice of $a=\psi(0)$ and $A_0$.

Such solutions stay regular as long as $\mu>0$.
Their global behaviour is characterized by three possibilities:

\begin{enumerate}
\item

$\mu$ has a zero for finite $r=r_0$ with diverging $\psi$, while $A,\baf$
and $\sqrt{\mu}\psi$ stay finite.
This type of singularity is just a coordinate singularity related to the
stationarity of $r$ considered as a metrical function.
In terms of a `geodesic' coordinate $\tau=\int^\tau_0\ov{dr}{\sqrt{\mu}}$
the function $r(\tau)$ reaches a maximum $r_0=r(\tau_0)$ and then runs back
to $r=0$ developing a curvature singularity there.
This type of behaviour is generic for solutions with a regular origin.

\item

$\mu$ has a zero for finite $r=\rh$ with finite $\psi,\baf$ and $A$.
Such solutions possess a cosmological horizon at $r=\rh$ enforcing 
the boundary conditions    

\bea\label{bdhor}
\mh'&=&\ov1{\rh}\Bigl[1-2\fh^2-\ov{\rh^2}4\bigl
   (\fh^2-\bet^2\bigr)^2\Bigr]\\                  
\ph&=&\ov{\fh}{\rh^2\mh'}\Bigl[2+\ov{\rh^2}2\bigl(\fh^2-
\bet^2\bigr)\Bigr]
\eea
with $\fh=\baf(\rh)$ etc..

\item

$\mu>0$ for all $r>0$, i.e.\ solutions staying finite for all $r$. The
asymptotic behaviour for $r\to\infty$ is given by
\be\label{bdinf1}
\mu(r)=1-2\bet^2+\ov{d}{r}+O(\ov1{r^2})
\ee
requiring $\bet<\ov1{\sqrt{2}}$
\be\label{bdinf2}
\baf(r)=\bet-\ov2{\bet r^2}+O(\ov1{r^3})+c\ov{e^{-\omega(r-
            \ov{d}{1-2\bet^2}\ln r)}}r\bigl(1+O(\ov1{r})\bigr)
\ee
with $\omega=\ov{\bet}{\sqrt{1-2\bet^2}}$ and some constants 
$c$ and $d$ depending on the solution.

The asymptotic behaviour of $\mu$ entails a `conical' singularity at
$r=\infty$ corresponding to deficit solid angle $\Delta=8\pi\bet^2$.
For $\bet^2=\ov1{2}$ this deficit angle becomes $4\pi$ and the character of
the solution for $r\to\infty$ changes to
\bea\label{asympt}
\mu(r)&=&\ov{d}{r}\bigl(1+O(\ov1{r})\bigr)\\                 
\baf(r)=&=&\ov1{\sqrt{2}}-\ov{2\sqrt{2}}{r^2}+O(\ov1{r^3})+
        ce^{-\omega r^{3/2}}\bigl(1+O(\ov1{r})\bigr)
\eea
with $\omega=\ov1{3}\sqrt{\ov2{d}}$ and again constants $c$ and $d$
depending on the solution.

For $\bet>1/\sqrt{2}$ the asymptotic behavior for $r\to\infty$ becomes
oscillatory due to the sign change of $\omega^2$. However, the factor $1/r$ 
in front of the exponential leads to a damping of the amplitude and
$\baf(r)\to\bet$ for $r\to\infty$ as is clearly visible from the numerical
results of \cite{Liebling}.

\end{enumerate}

Both the solutions of type 2) and 3) require fine-tuning of the parameter
$a$ characterizing solutions regular at $r=0$.
The numerical results of \cite{Liebling} indicate that for $0<\bet<\betam$ 
exactly one such solution with non-vanishing Higgs field exists.
The corresponding graph of $a(\bet)$ is shown in Fig.\ref{figa}.
In the following we shall try to explain the essential features of
$a(\bet)$.
\begin{figure}   
\hbox to\hsize{\hss
   \epsfig{file=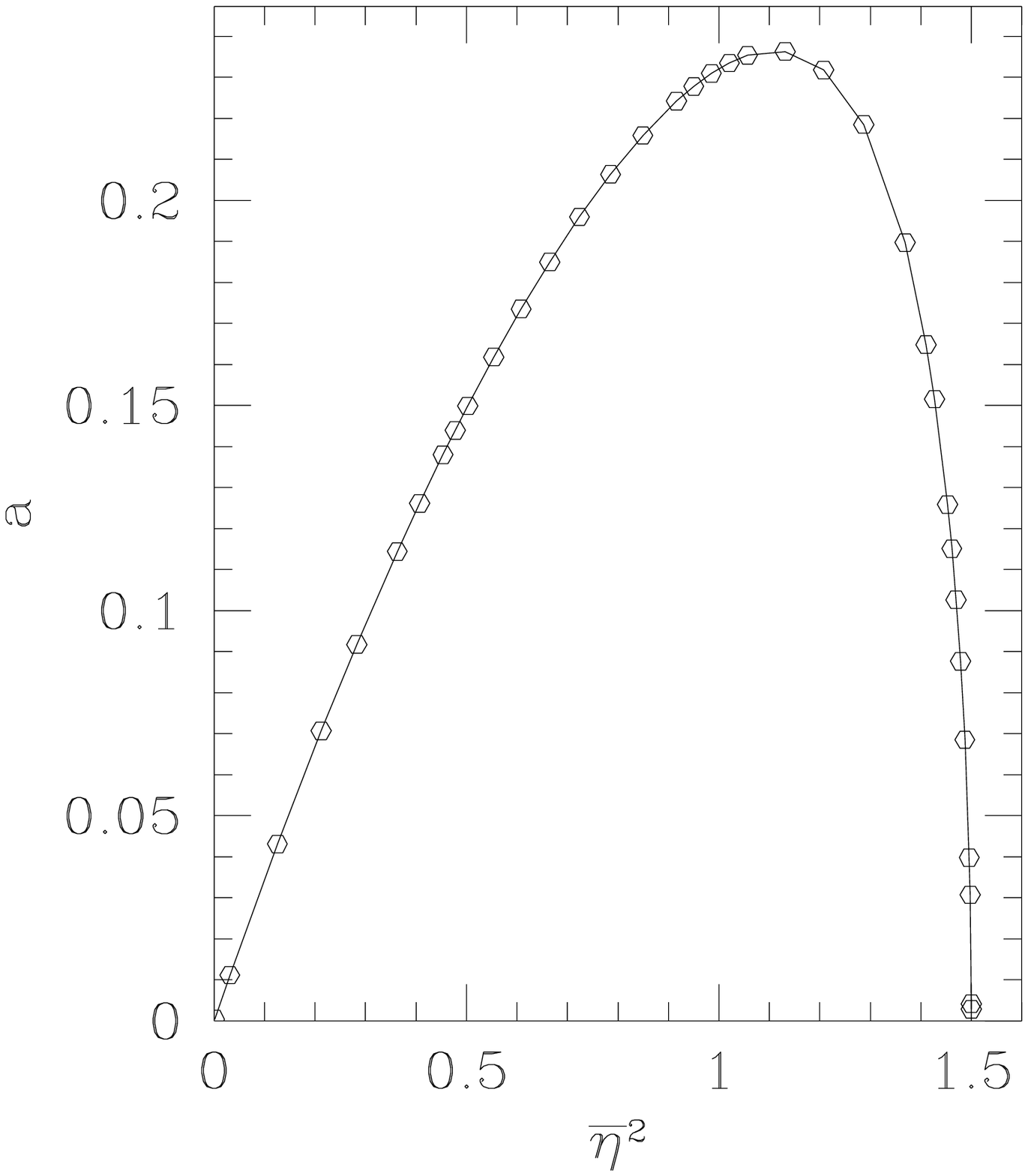,width=\hsize,%
   bbllx=0.5cm,bblly=4.5cm,bburx=20.0cm,bbury=21cm}%
  }
\labelcaption{figa}{
$a=\baf'(0)$ for the static monopole as a function of $\bet^2$.
Observe that $a\rightarrow 0$ as $\bet^2 \rightarrow 3/2$
signifying the approach to de Sitter
space. Note that the critical point in $a$ does not occur
at $\bet^2 = 1$, which was found to be the critical point in the
core radius in~\cite{Liebling}.
}
\end{figure}

For small values of $\bet$ the self-gravitating monopole can be understood
as a perturbation of the flat monopole, which is obtained for 
$\bet\to 0$  after a rescaling $\baf=\bet\hat f$ 
and $r=\hat r/\bet$ solving the equation
\be\label{flat}
\bigl(\hat r^2\hat f'\bigr)'=\hat f\Bigl[2+\ov{\hat r^2}2\bigl(
       \hat f^2-1\bigr)\Bigr]\,. 
\ee
Solving this equation numerically with the boundary condition 
$\hat f(\hat r)=\hat a\hat r+O(\hat r^3)$ one finds a globally regular
solution for $\hat a\approx 0,3578$. 
Expressed in terms of $\baf$ we get
\be\label{small}
\baf(r)=\bet^2\hat a r+O(r^3)=ar+O(r^3)
\ee
and thus $a=\bet^2\hat a$.
%
%
This behavior is clearly visible in Fig.\ref{figa}.

We also see from this figure that the function $a(\bet)$ after reaching a
maximum turns back to $a=0$ with diverging derivative. The latter may be
understood as a consequence of the invariance $\baf\to -\baf$ of the field
equations in combination with the smooth behaviour of $a(\bet)$ at $\betam$.
Next we will use this ``empirical'' observation to 
derive the exact value for $\betam$.

Suppose we take a globally regular solution $\baf(a(\bet),\bet,r)$ for some
fixed value of $\bet$ and vary $a(\bet)\to a(\bet)+\delta a$. 
Then generically we will get a solution of type 1). However, if $\partial
a(\bet)/\partial\bet=\infty$ an infinitesimal change of $a$ will carry us to
another regular solution for the same value of $\bet$. 
Thus $\partial \baf/\partial a$ will be bounded for $r\to\infty$ resp.
$r\to\rh$ for points where $\partial a/\partial\bet=\infty$ 
in contrast to points where this derivative is  finite.
As seen from Fig.\ref{figa} the solution for $\bet=\betam$ corresponds
to $a(\bet)=0$ and thus $\baf(\betam,r)\equiv 0$.
This is nothing but the de Sitter solution given by integrating Eq.(\ref{fequs_mu})
\be\label{deSit}
\mu_{\rm dS}(r)=1-\Bigl(\ov{r}{\rh}\Bigr)^2 \quad {\rm with}\quad 
     \rh=\ov{2\sqrt{3}}{\bet^2}\,.
\ee
Figure~\ref{figdS} shows the approach of the monopole solutions to the
de Sitter solution for $\bet\to\betam$.
\begin{figure}   
\hbox to\hsize{\hss
   \epsfig{file=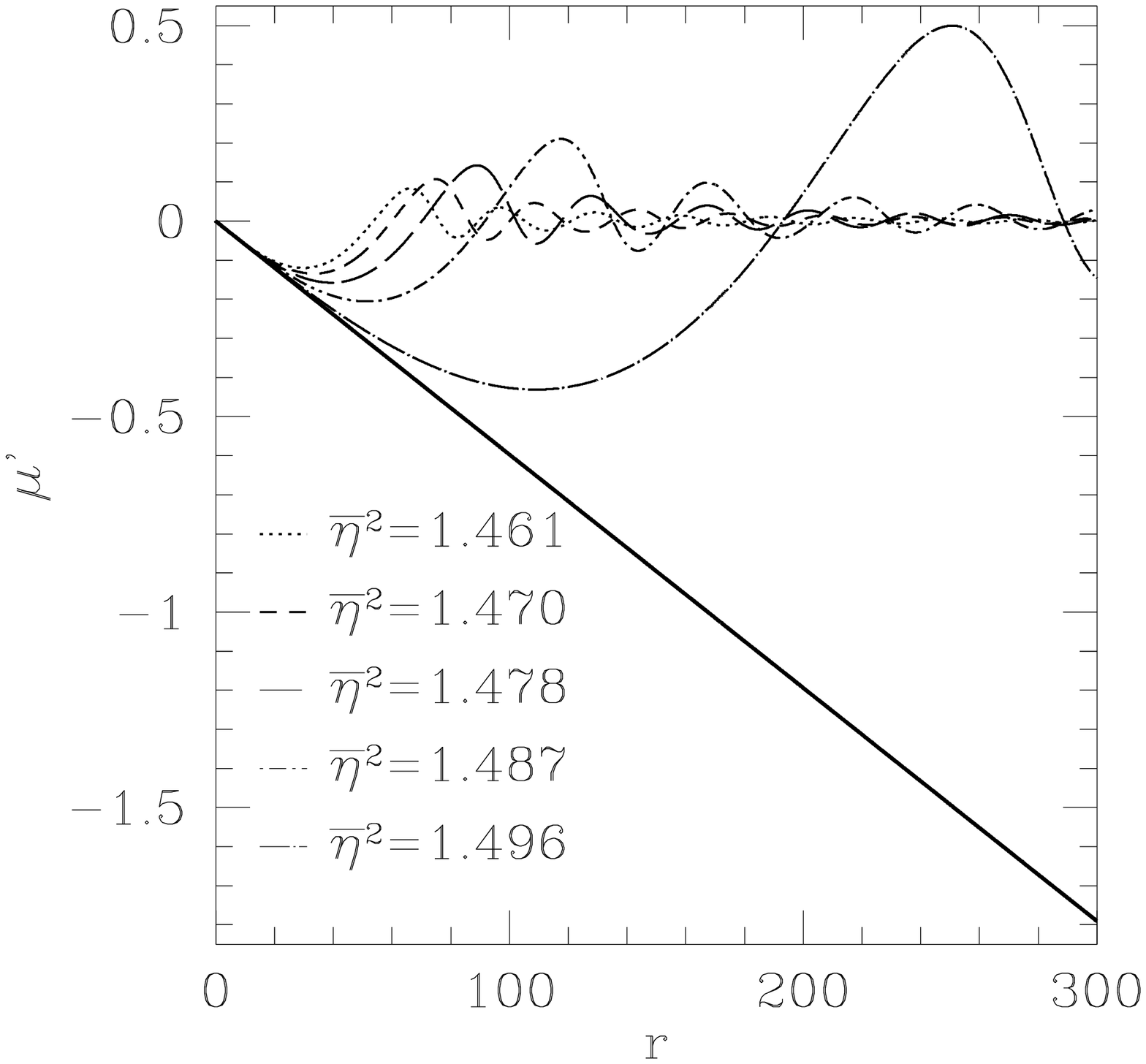,width=\hsize,%
   bbllx=0.5cm,bblly=5.5cm,bburx=19.0cm,bbury=22cm}%
  }
\labelcaption{figdS}{
Demonstration of approach to de Sitter solution. The field $\mu'(r)$
is shown as $\bet^2$ approaches $3/2$. The solid line denotes
the de Sitter solution for $\bet^2=3/2$
while the other profiles show $\mu'$ for
the static monopoles. The solutions approach the de Sitter solution for an
increasing domain of $r$ until ultimately driven to oscillate around
the regular solution $\mu'=0$.
}
\end{figure}
According to what was said above, the de Sitter solution must have the bounded
``zero mode'' $\partial\baf/\partial a$ for $\bet=\betam$.
Linearizing the field Eqs.(\ref{fequs_f}-\ref{fequs_a}) on the 
de Sitter background we obtain (putting $\varphi=r\delta\baf$) 
\be\label{lin1}
\bigl(\mu_{\rm dS}\varphi'\bigr)'\equiv\Bigl(\bigl(1-
          \ov{r^2}{\rh^2}\bigr)\varphi'\Bigr)'=
          \Bigl(\ov2{r^2}-\ov{\bet^4}6-\ov{\bet^2}2\Bigr)\varphi\,.
\ee
Introducing $x=r/\rh$ we get
\be\label{lin2}
\frac{d}{dx}\Bigl(\bigl(1-x^2\bigr)\frac{d}{dx}\varphi\Bigr)=
     \Bigl(\ov2{x^2}-2-\ov6{\bet^2}\Bigr)\varphi\,.
\ee
The solution with the correct boundary condition $\varphi(x)=x^2+O(x^4)$
for $x\to 0$ valid for $\varphi=\partial\baf/\partial a$ is given
by $\varphi\equiv x^2$ obtained for $\bet^2=3/2$.

Thus we have found a simple derivation of $\betam=\sqrt{\ov{3}{2}}$.
Since the r.h.s. of Eq.(\ref{lin2}) becomes more negative for $\bet^2<3/2$
the solution $\varphi(x)$ fulfilling the correct boundary condition
at $x=0$ develops a zero before reaching the horizon at $x=1$.
According to the ``Jacobi criterion'' \cite{Gelfand} this implies an
instability of the de Sitter background for $\bet<\sqrt{\ov3{2}}$.
Linearizing the time-dependent field equations around the static solution
we let
\be\label{time}
\baf(r,t)=\baf_0(r)+e^{i\omega t}\delta \baf,
\ee
where $\baf_0$ represents the static Higgs field.
Unstable modes of the static solution correspond to modes with 
$\omega^2<0$.
The Jacobi criterion relates the existence of such negative modes
to the existence of zeros of the solution for $\omega^2=0$ (zero mode) 
with the correct boundary condition at $r=0$. 

For $\bet>\sqrt{\ov3{2}}$ the solution $\varphi(x)$ remains
positive for all $x$ indicating the stability of the solution under
spherically symmetric perturbations with support inside the horizon.
On the other hand for $\bet\to 0$ the de Sitter solution 
develops more and more unstable modes
manifested by the existence of bounded zero-modes with more and more zeros.
The latter are readily obtained with the polynomial ansatz
\be\label{zero}
\varphi_K(x)=\sum^K_{k=1}c_k x^{2k}\,.
\ee                                
Inserting this ansatz into the Eq.(\ref{lin2}) yields the recursion
relation 
\be\label{recur}
c_{k+1}=\ov{2k(2k+1)-2-\ov6{\bet^2}}{(2k+2)(2k+1)-2}\,c_k
\ee                                
and the condition (derived from $c_{K+1}=0$)
\be\label{eta}
\bet_K^2=\ov{3}{K(2K+1)-1}
\ee                                
yielding $\bet^2_1=3/2$, $\bet^2_2=1/3$, $\bet^2_3=3/20$ etc..
\begin{figure}   
\hbox to\hsize{\hss
   \epsfig{file=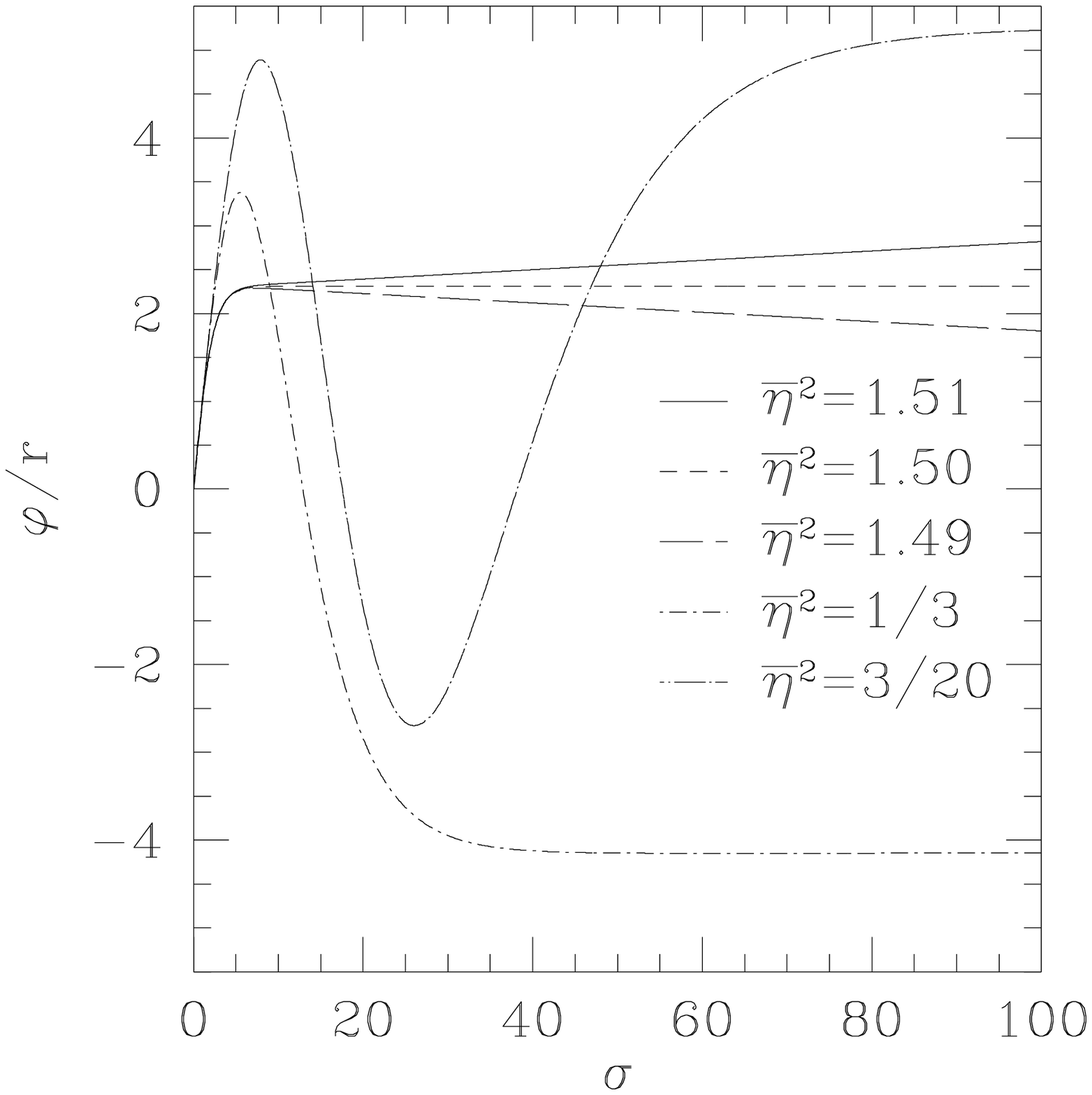,width=\hsize,%
   bbllx=0.5cm,bblly=4.5cm,bburx=20.0cm,bbury=21cm}%
  }
\labelcaption{figy}{Perturbations to the de Sitter solution.
For $\bet^2 > 1.5$, the perturbation is everywhere positive
indicating linear stability. For $\bet^2 < 1.5$, zero-crossings exist
indicating instability of the de Sitter solution. Additional
zeros appear as $\bet^2$ passes through the values $\bet_K^2$ found
from Eq.(\ref{eta}), the first few of which are shown here.
}
\end{figure}

Eq.(\ref{lin2}) can be transformed to the hypergeometric differential
equation as follows.
Putting $\xi=\ov{1+x^2}{1-x^2}$ and $\varphi(x)=\xi(1+\xi)^my(\xi)$ with
$2m(2m-1)=3+6/\bet^2$ we obtain  
\be\label{hyper}
\xi(\xi+1)y''+\Bigl(\bigl(2m+3)\xi+\ov5{2}\Bigr)y'+(m+1)^2y=0
\ee                                
fulfilled by the hypergeometric function $F(m+1,m+1,5/2,\xi)$
\cite{Bateman},
the polynomial solutions $\varphi_K(x)$ corresponding to the Jacobi polynomials
$y_K(\xi)$.

The stability of the flat global monopole (under spherically symmetric
perturbations) implies its stability for small
values of $\bet$. According to standard arguments for one-parameter families
of solutions \cite{Maeda}
we expect no change of stability up to the bifurcation point 
$\bet=\betam$ with the de Sitter solution.
In order to analyze this hypothesis we have performed a detailed numerical 
stability analysis for the global monopole using again the Jacobi criterion.
For solutions with a horizon our stability analysis is restricted to the
region inside the horizon. This is justified by the fact that no
perturbations can penetrate through the horizon from the region where
$\mu<0$ (similar to the case of stability analysis performed for black
holes).
\begin{figure}   
\hbox to\hsize{\hss
   \epsfig{file=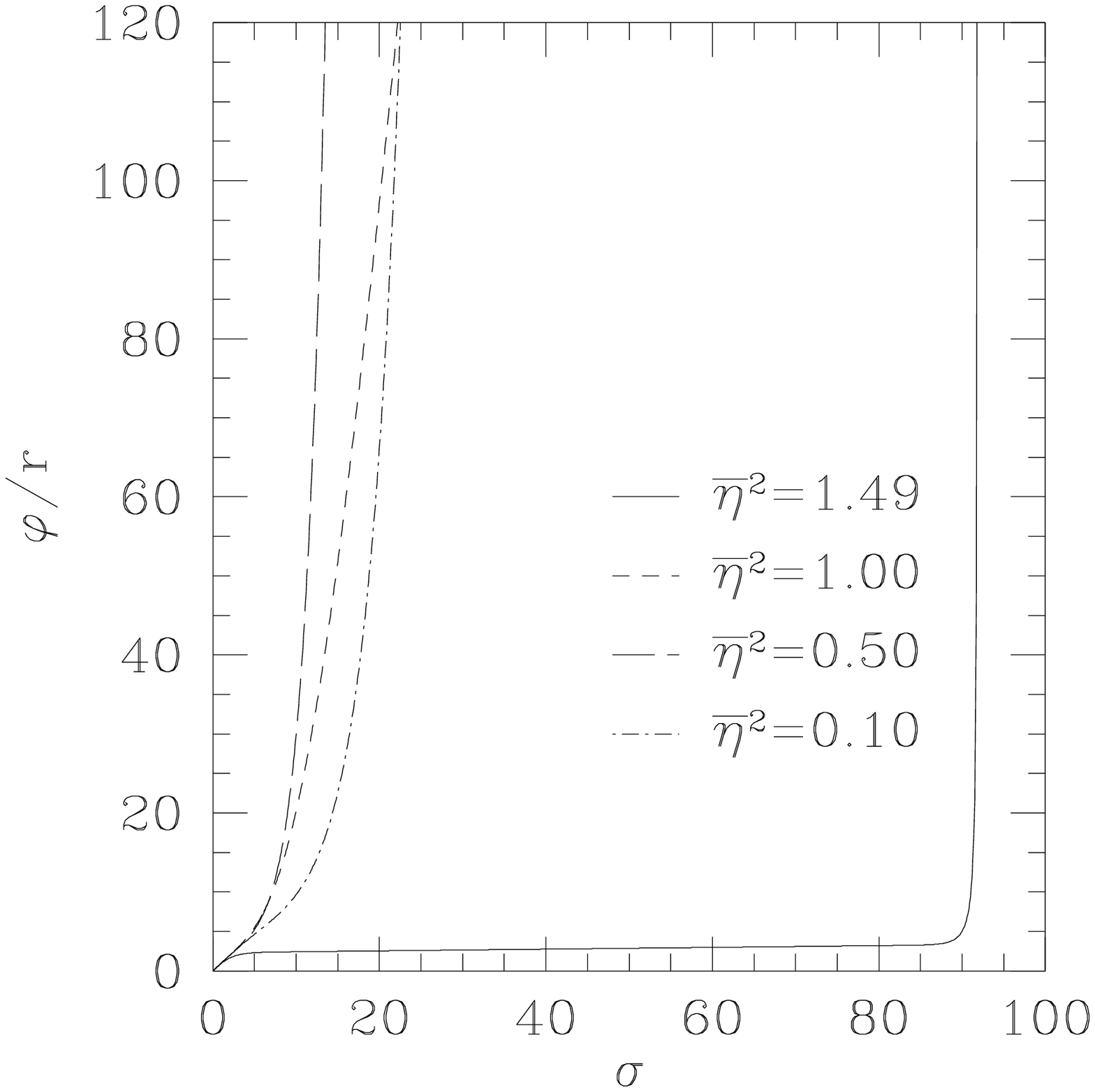,width=\hsize,%
   bbllx=0.5cm,bblly=4.5cm,bburx=20.0cm,bbury=21cm}%
  }
\labelcaption{figst}{Perturbations to the static global monopole.
For all $\bet^2$, the perturbation is everywhere positive
indicating linear stability.
}
\end{figure}

As is well-known \cite{Hollmann} the metric perturbations can be integrated
out and one obtains a Schr\"odinger type perturbation equation for
$\varphi$.
Introducing the coordinate $\sigma$ through
$\sigma=\int^\sigma_0\ov{dr}{A\mu}$ one gets
\be\label{Schroed}
-\ov{d^2\varphi}{d\sigma^2}+V\varphi=0
\ee                                
with 
\be\label{pot}
V=A^2\mu\Bigl[\ov2{r^2}+\ov1{2}(3\baf^2-\bet^2)+\ov2{r^2A}(A\mu r^3\psi^2)'
  +\ov{(A\mu)'}{rA}\Bigr]
\ee                                
and the boundary condition $\varphi(r)=\sigma^2+O(\sigma^4)$ for $\sigma\to 0$.
Our numerical analysis shows (see Fig.\ref{figst})
that the solutions $\varphi$ have no zero for all
$0<\bet<\sqrt{3/2}$ confirming our hypothesis about the stability of the
global monopole.

\section{Acknowledgments}
This work began at the ITP at the
University of California, Santa Barbara on occasion of the workshop on
Classical and Quantum Physics of Strong Gravitational Fields
and has been supported in part by the National Science Foundation
under Grant No. PHY94-07194.
The hospitality of the ITP is gratefully acknowledged.
SL is also thankful for the financial support of Southampton College.


\begin{thebibliography}{10}

\newcommand{\NPB}{\sl Nucl.\ Phys.\ \bf B\,}
\newcommand{\PLB}{\sl Phys.\ Lett.\ \bf B\,}
\newcommand{\PRD}{\sl Phys.\ Rev.\ \bf D\,}
\newcommand{\PRL}{\sl Phys.\ Rev.\ Lett.\ \bf}

\bibitem{Liebling}
S.L. Liebling,
``Static gravitational global monopoles'',\\
gr-qc/9906014.

\bibitem{Linde}
A. Linde,
{\PLB 327} (1994) 208.

\bibitem{Vilenkin}
A. Vilenkin,
{\PRL 72} (1994) 3137.

\bibitem{LieblingC}
S.L. Liebling,
{\PRD 60} 061502 (1999).

\bibitem{Gelfand}
I.M. Gelfand and S.V. Fomin,
{\sl Calculus of Variations},\\
Englewood Cliffs, N.J., Prentice Hall, 1963.

\bibitem{Bateman}
A.~Erd\'elyi, M.~Magnus, F.~Oberhettinger and F.G.~Tricomi,
{\sl Higher Transcendental Functions, Vol.2},\\
McGraw Hill, New York, 1953.

\bibitem{Maeda}
K. Maeda, T. Tachizawa, T. Torii and T. Maki,
{\PRL 72} (1994) 450.

\bibitem{Hollmann}
H. Hollmann,
{\PLB 338} (1994) 181.

\end{thebibliography}
\end{document}